# An X-Ray Regenerative Amplifier Free-Electron Laser Using Diamond Pinhole Mirrors

H.P. Freund,[1,2] P.J.M. van der Slot,[3,2] and Yu. Shvyd'ko[4]
[1] Department of Electrical and Computer Engineering, University of New Mexico, Albuquerque, NM, USA
[2]Calabazas Creek Research, San Mateo, California, USA
[3]Mesa+ Institute for Nanotechnology, University of Twente, Enschede, the Netherlands
[4]Advanced Photon Source, Argonne National Laboratory, Argonne, Illinois, USA

Free-electron lasers (FELs) have been built ranging in wavelength from long-wavelength oscillators using partial wave guiding through ultraviolet through hard x-ray FELs that are either seeded or start from noise (SASE). Operation in the x-ray spectrum has relied on single-pass SASE due either to the lack of seed lasers or difficulties in the design of x-ray mirrors. However, recent developments in the production of diamond crystal Bragg reflectors point the way to the design of regenerative amplifiers (RAFELs) which are, essentially, low-*Q* x-ray free-electron laser oscillators (XFELOs) that out-couple a large fraction of the optical power on each pass. A RAFEL using a six-mirror resonator providing out-coupling of 90% or more through a pinhole in the first downstream mirror is proposed and analyzed using the MINERVA simulation code for the undulator interaction and the Optics Propagation Code (OPC) for the resonator. MINERVA/OPC has been used in the past to simulate infrared FEL oscillators. For the present purpose, OPC has been modified to treat Bragg reflection from diamond crystal mirrors. The six-mirror resonator design has been analyzed within the context of the LCLS-II beamline under construction at the Stanford Linear Accelerator Center and using the HXR undulator which is also to be installed on the LCLS-II beamline. Simulations have been run to optimize and characterize the properties of the RAFEL, and indicate that substantial powers are possible at the fundamental (3.05 keV) and third harmonic (9.15 keV).

## I. INTRODUCTION

Development of x-ray free electron lasers (XFELs) began in the United States with the proposal for the Linac Coherent Light Source (LCLS) at the Stanford Linear Accelerator Center (SLAC) culminating with the first lasing in 2009 [1]. The success of the LCLS encouraged the development of other XFELs worldwide [2-5]. Due to the lack of seed lasers at x-ray wavelengths, however, each of these facilities rely upon Self-Amplified Spontaneous Emission (SASE) in which the optical field grows from shot noise on the electron beam to saturation in a single pass through a long undulator. While pulse energies of the order of 2 milliJoules have been achieved at Ångstrom to sub-Ångstrom wavelengths, SASE exhibits shot-to-shot fluctuations in the output spectra and power of about 10 – 20 percent. For many applications, these fluctuations are undesirable, and efforts are underway to find alternatives to SASE.

One alternative relies upon a High-Gain Harmonic Generation (HGHG) cascade. In HGHG [6,7] two undulator sections are used. In the first section, called the *Modulator*, the electron beam is injected in conjunction with the output from a high-power seed laser that imposes a correlated velocity modulation on the electrons. The modulated electrons then pass through a magnetic dispersive section (often a magnetic dipole chicane) which efficiently induces a substantial density modulation, and then enters the second undulator section, called the *Radiator*, which is tuned to a harmonic of the resonance in the *Modulator*. Since the electrons have been preconditioned for emission in the *Radiator*, the harmonic power grows coherently and rapidly to saturation. It is possible to generate successively shorter wavelengths by cascading multiple HGHG segments as demonstrated in the FERMI facility in Trieste, Italy [8]. However, at the present time, no HGHG cascade has achieved x-ray wavelengths.

Another alternative is represented by schemes for the self-seeding of SASE [9-13]. Here, SASE is terminated prior to saturation by a break in the undulator, and the optical pulse is filtered by a monochromator. The filtered pulse is reintroduced to the electrons in the remainder of the undulator where the desired wavelength is amplified. Self-seeding is useful when the over-riding concern is narrow bandwidth. However, it is subject to large intra-pulse fluctuations in SASE and shot-to-shot fluctuations in the electron beam from the linac. As a result, self-seeding is not a universal solution to the fluctuations associated with SASE.

The utility of an x-ray FEL oscillator (XFELO) has been under study for a decade [14-21] making use of resonators based upon Bragg scattering from high-reflectivity diamond crystals [22-26]. The development of these crystals is a major breakthrough in the path toward an XFELO. Estimates indicate that using a superconducting rf linac producing 8 GeV electrons at a 1 MHz repetition rate is capable of producing $10^{10}$ photons per pulse at a 0.86 Å wavelength with a FWHM bandwidth of about $2.1 \times 10^{-7}$. This design is consistent with the LCLS-II High Energy Upgrade [27]. As a consequence, an XFELO on a facility such as the LCLS-II and LCLS-II-HE can be expected to result in a decrease in SASE fluctuations in the power and spectrum and to narrow the spectral linewidth.

As with the majority of FELOs to date, such as the IR Demo [28,29] and 10-kW Upgrade [30] experiments at Thomas Jefferson National Accelerator Facility, the aforementioned XFELOs employ low gain/high-*Q* resonators with transmissive out-coupling through thin diamond crystals [21]. Potential difficulties with low-gain/high-$Q$ resonators derive from sensitivities to electron beam and mirror loading and alignments. For example, the Jefferson Lab 10-kW Upgrade had large fluctuations at high power from thermal mirror loading. In addition, transmissive out-coupling with the high intra-cavity powers associated with high-$Q$ resonators can result in mirror damage. While experiments show that diamond crystals can sustain relatively high thermal and radiation loads [24], transmissive out-coupling cannot be easily achieved at the photon energies of interest in the present paper. Hence, we consider an XFELO design with high out-coupling efficiency using a pinhole diamond mirror based on a Regenerative Amplifier (RAFEL) concept [31,32].

The RAFEL is based on a high-gain/low-$Q$ resonator where the majority of the power is out-coupled on each pass. Typically, this ranges from 90 – 95% of the power. As a result, the power loading on the mirrors is significantly reduced relative to that in a high-$Q$ resonator. Since the interaction in a RAFEL optically guides the light, the optical mode is characterized by high purity with $M^2 \approx 1$ whether hole or transmissive out-coupling is used [33]. This might include an unstable resonator; however, it was shown by Siegman [34] that gain guiding, such as in an FEL, will stabilize a resonator that is otherwise (*i.e.*, *in vacuo*) unstable.

RAFELs differ from low-gain FELOs in some respects and resemble SASE FELs in others. One difference is that, as in Madey's theorem, a low-gain oscillator has no gain directly on resonance. In contrast, the growth in the exponential regime has a peak on-resonance, and this affects the wavelengths excited in a RAFEL. A second difference is the efficiency. The efficiency, $\eta$, of a low-gain oscillator is $\eta \approx 1/2N_w$, for $N_w$ undulator periods [35]. Since the radiation exponentiates in each pass through the undulator in a RAFEL, the efficiency is given by that for the high-gain Compton regime where $\eta \approx \rho$ (*i.e.*, the Pierce parameter). For the Jefferson Lab 10-kW Upgrade, $N_w = 28$ (with one additional period in the entry/exit transitions) and $\eta \approx 1.8\%$. However, even a low-gain XFELO requires a much longer undulator with a correspondingly lower efficiency; hence, an efficiency dependent on the Pierce parameter presents no obstacle to an x-ray RAFEL. A third difference is the linewidth, which scales inversely with $N_w$ in a low-gain oscillator but is given by $\Delta\omega/\omega \approx \rho$ in the high-gain Compton regime (assuming that the cavity contains no additional filtering). A fourth difference is the transverse mode structure, which is determined largely by the resonator in a low-gain oscillator. In a RAFEL, by contrast, exponential gain optically guides the radiation and the mode structure is strongly governed by the undulator. A fifth difference is in the effect of slippage. Slippage in a low-gain oscillator scales with $N_w$. However, the high-gain in a RAFEL results in a reduction in the group velocity [36] and in slippage which scales $\propto N_w/3$. One similarity that the RAFEL shares with low-gain oscillators is the presence of limit-cycle oscillations [33,37].

In this paper, we analyse an x-ray RAFEL configuration based upon a six-mirror resonator composed of diamond crystal Bragg reflectors. High-efficiency out-coupling is achieved through a pinhole in one of the diamond crystals. We consider that the resonator is implemented on the LCLS-II beamline at SLAC using the HXR (*i.e.*, high energy x-ray) undulator and produces x-ray photons at energies of 3.05 keV in the fundamental and 9.15 keV at the third harmonic. Simulations of the RAFEL are conducted using the MINERVA simulation code for the undulator interaction and the Optics Propagation Code (OPC) to describe the propagation of the x-rays through the resonator.

The organization of the paper is as follows. The general properties of MINERVA and OPC are described in Sec. II. The resonator is described in Sec. III, and the simulations of the RAFEL are discussed in Sec. IV. A summary and discussion are given in Sec. V.

## II. GENERAL SIMULATION PROPERTIES

The MINERVA formulation [38,39] describes the particles and fields in three spatial dimensions and includes time dependence as well. Electron trajectories are integrated using the complete Newton-Lorentz force equations. No wiggler-averaged-orbit approximation is made. The magnetostatic fields can be specified by analytical functions for a variety of analytic undulator models (such as planar, elliptical, or helical representations), quadrupoles, and dipoles. These magnetic field elements can be placed in arbitrary sequences to specify a variety of different transport lines. As such, field configurations can be described for single or multiple wiggler segments with quadrupoles either placed between the undulators or superimposed upon the undulators to create a FODO lattice. Dipole chicanes can also be placed between the undulators to model various optical klystron and/or high-gain harmonic generation (HGHG) configurations. The fields can also be imported from a field map.

The electromagnetic field is described by a modal expansion. For free-space propagation, we use Gaussian optical modes. The Gauss-Hermite modes are used for simulation of planar undulators, while Gauss-Laguerre modes are used for elliptical or helical undulators.

The electromagnetic field representations are also used in integrating the electron trajectories, so that harmonic motions and interactions are included in a self-consistent way. Further, the same integration engine is used within the undulator(s) as in the gaps, quadrupoles, and dipoles, so that the phase of the optical field relative to the electrons is determined self-consistently when propagating the particles and fields in the gaps between the undulators.

MINERVA has also been linked to the Optics Propagation Code (OPC) [40,41] for the simulation of FEL oscillators or propagating an optical field beyond the end of the undulator line to a point of interest. OPC propagates the optical field using either the Fresnel diffraction integral or the spectral method in the paraxial approximation using fast discrete Fourier transforms (FFT). A modified Fresnel diffraction integral [34, 42] is also available and allows the use of FFTs in combination with an expanding grid on which the optical field is defined. This method is often used when diffraction of the optical field is large. Propagation can be done either in the time or frequency domain. The latter allows for the inclusion of dispersion and wavelength dependent properties of optical components. Currently, OPC includes mirrors, lenses, phase and amplitude masks, and round and rectangular diaphragms. Several optical elements can be combined to form more complex optical component, *e.g.*, by combining a mirror with a hole element, extraction of radiation from a resonator through a hole in one of the mirrors can be modelled. Phase masks can be used, for example, to model mirror distortions or to create non-standard optical components like a cylindrical lens.

In a typical resonator configuration, OPC handles the propagation from the end of the gain medium to the first optical element, applies the action of the optical element to the optical field and propagates it to the next optical element and so on until it reaches the entrance of the gain medium. Diagnostics can be performed at the planes where the optical field is evaluated. Some optical elements, specifically diaphragms and mirrors allow forking of the optical path. For example, the reflected beam of a partial transmitting output mirror forms the main intracavity optical path, while the transmitted beam is extracted from the resonator. When the intracavity propagation reaches the output mirror, this optical propagation can be temporarily suspended, and the extracted beam can be propagated to a diagnostic point for evaluation. Then the intra-cavity propagation (main path) is resumed.

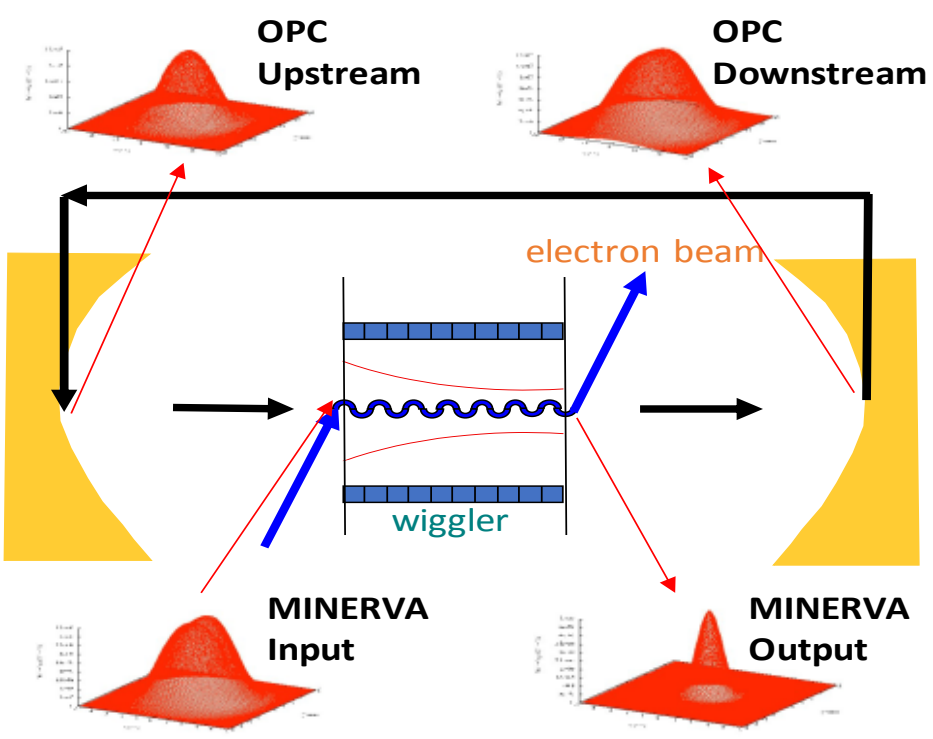

Fig. 1: Schematic illustration of the operation of MINERVA/OPC for a concentric resonator.

The numerical procedure involves translating between the input/output required for MINERVA and OPC. The procedure is illustrated schematically for a concentric resonator in Fig. 1. Initially, we run the FEL simulation to determine the optical output after the first pass through the undulator, which then writes a file describing the complex field of the optical mode. OPC is then used to propagate this field to the downstream mirror, which is partially transmissive in the current example. The portion of the optical mode that is reflected is then propagated to the upstream mirror (which is a high reflector) by OPC, and then back to the undulator entrance. The field at the undulator entrance is then reduced to an ensemble of Gaussian modes that is used as input to the FEL simulation for the next pass. This process is repeated for an arbitrary number of passes.

OPC has been modified to be able to treat the reflections from the diamond crystal Bragg mirrors where the mirror losses and angles of reflection depends on the crystal orientation/geometry and the x-ray energy/wavelength. X-ray Bragg mirrors typically have a very narrow reflection bandwidth and a very narrow angle of acceptance [43]. To correctly implement the action of such Bragg reflectors, OPC first calculates a temporal Fourier transform. For the RAFEL under consideration and for computational efficiency, this is done once at the beginning of the optical path when the optical field is handed over from MINERVA to OPC and the propagation is performed in the wavelength domain, *i.e.*, each wavelength is independently propagated along the optical path through the resonator. At the end of the optical path, before the optical is handed over to MINERVA, the inverse temporal Fourier transform is calculated. As the optical field inside the cavity is typically not collimated, a spatial Fourier transform in the transverse coordinates is calculated for each of the wavelengths when a Bragg mirror is encountered along the optical path. Each combination of transverse and longitudinal wavenumber corresponds to a certain photon energy and angle of incidence on the Bragg mirror and these parameters are used to calculate the complex reflection and transmission coefficients of the Bragg mirror [43]. After applying the appropriate parameter to the optical field, depending on whether it is reflected or transmitted, the inverse spatial Fourier transform is calculated and the optical field is propagated to the next optical element along the optical path until the end is reached.

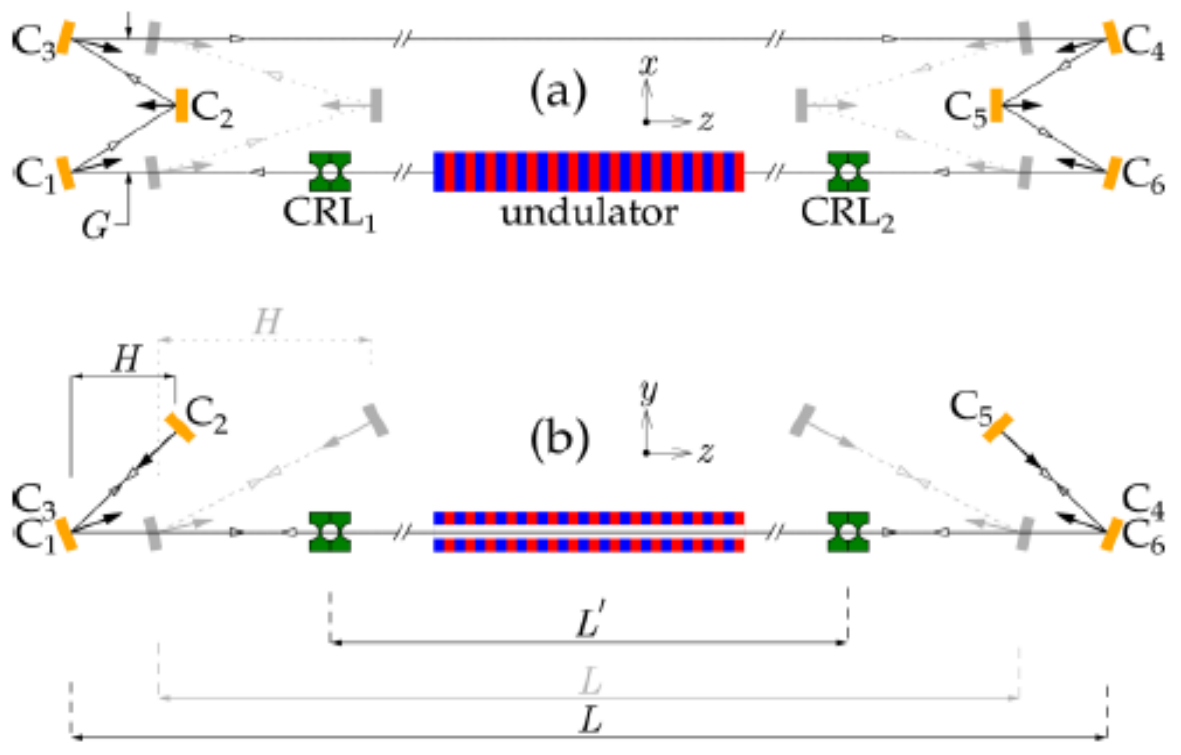

Fig. 2: Tunable, six-crystal resonator [25].

We chose to operate on a resonance with 3.05 keV photons (4.07 Å wavelength) using an on-axis undulator field of 5.605 kG, and the angles of reflection $\theta = 81.3°$ for the six diamond crystal mirrors set into the (111) Bragg diffraction in the resonator.

MINERVA/OPC has been validated by comparison with the 10 kW Upgrade experiment at Jefferson Laboratory [44] and has also been used to simulate a RAFEL with a ring resonator [33]. Note that this means that OPC is capable of simulating FELOs at wavelengths from the infrared through the x-ray spectra.

## III. THE DIAMOND CRYSTAL BRAGG RESONATOR

We consider a six-crystal, tunable, compact cavity [25] which is illustrated in Fig. 2 [the top view is shown in (a) and the side view is shown in (b)]. The crystals are arranged in a non-coplanar (3-D) scattering geometry. There are two *backscattering* units comprising three crystals ($C_1$, $C_2$, and $C_3$) on one side of the undulator and three crystals ($C_4$, $C_5$, and $C_6$) on the other side. Collimating and focusing elements are shown as $CRL_{1,2}$, which could be grazing-incidence mirrors but are represented in the figure by another possible alternative – compound refractive lenses [45,46]. In each backscattering unit, three successive Bragg reflections take place from three individual crystals to reverse the direction of the beam from the undulator. Assuming that all the crystals and Bragg reflections are the same, the Bragg angles can be chosen within the range $30^\circ < \theta < 90^\circ$; however, Bragg angles close to $\theta = 45^\circ$ should be avoided to ensure high reflectivity for both linear polarization components, as the reflection plane orientations for each crystal change. The cavity allows for tuning the photon energy in a large spectral range by synchronously changing all Bragg angles. In addition, to ensure constant time of flight, the distance $L$ (which brackets the undulator), and the distance between crystals as characterized by $H$ have to be changed with θ. The lateral size $G$ is kept constant as the resonator is tuned.

Because the $C_1C_6$ and $C_3C_4$ lines are fixed, intra-cavity radiation can be out-coupled simultaneously for several users at different places in the cavity, although we only consider out-coupling through $C_6$ at the present time. Out-coupling through crystals $C_3$ and $C_6$ are most favourable, since the direction of the out-coupled beams do not change with photon energy, but out-coupling for more users through crystals $C_3$ and $C_6$ are also possible. Such multi-user capability is in stark contrast with present SASE beamlines which support one user at a time.

## IV. RAFEL SIMULATIONS

We consider the beamline associated with the LCLS-II [27]. This corresponds to an electron energy of 4.0 GeV, a bunch charge in the range of 10 – 30 pC with an rms bunch duration (length) at the undulator of 2.0 – 173 fs (0.6 – 52 μm) and a repetition rate of 1 MHz. The peak current at the undulator is 1000 A with a normalized emittance of 0.2 – 0.7 mm-mrad, and an rms energy spread of about 125 – 1500 keV.

The undulator that we consider in conjunction with this beamline is referred to as the HXR undulator (see Fig. 3) which is a plane-polarized, hybrid permanent magnet undulator with a variable gap, a period of 2.6 cm, and a peak magnetic field of 10 kG. Each HXR undulator is composed of 130 periods, and we consider that the first and last period describe an entry/exit taper. There is a total of 32 segments that can be installed. The break sections between the undulators are 1.0 m in length and contain quadrupoles, BPMs, etc., although we only consider the quadrupoles in the simulation which we consider to be located in the center of the break sections. The quadrupoles are assumed to be 7.4 cm in length with a field gradient of 1.71 kG/cm.

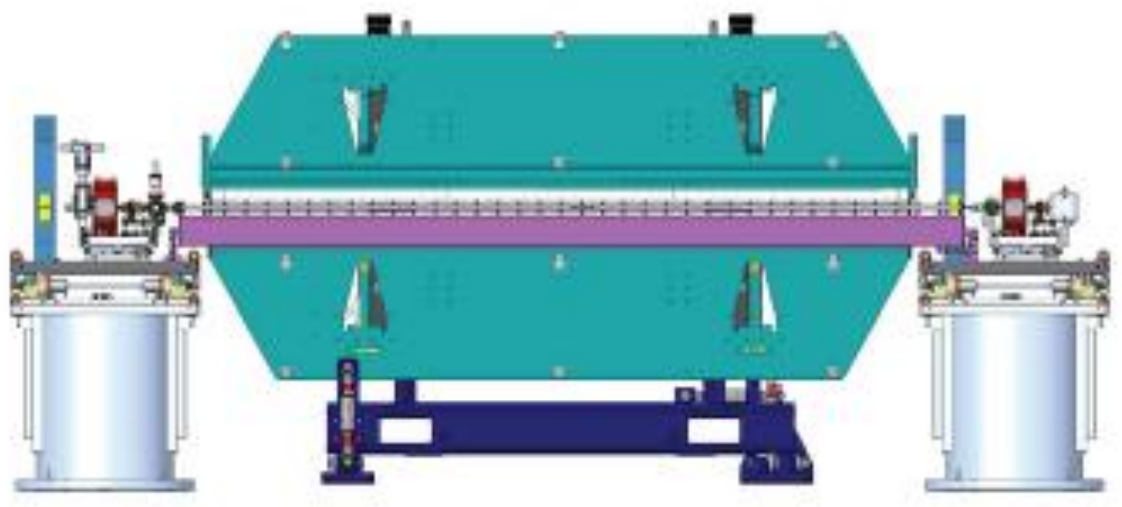

Fig. 3: Illustration of the LCLS-II HXR undulator [27].

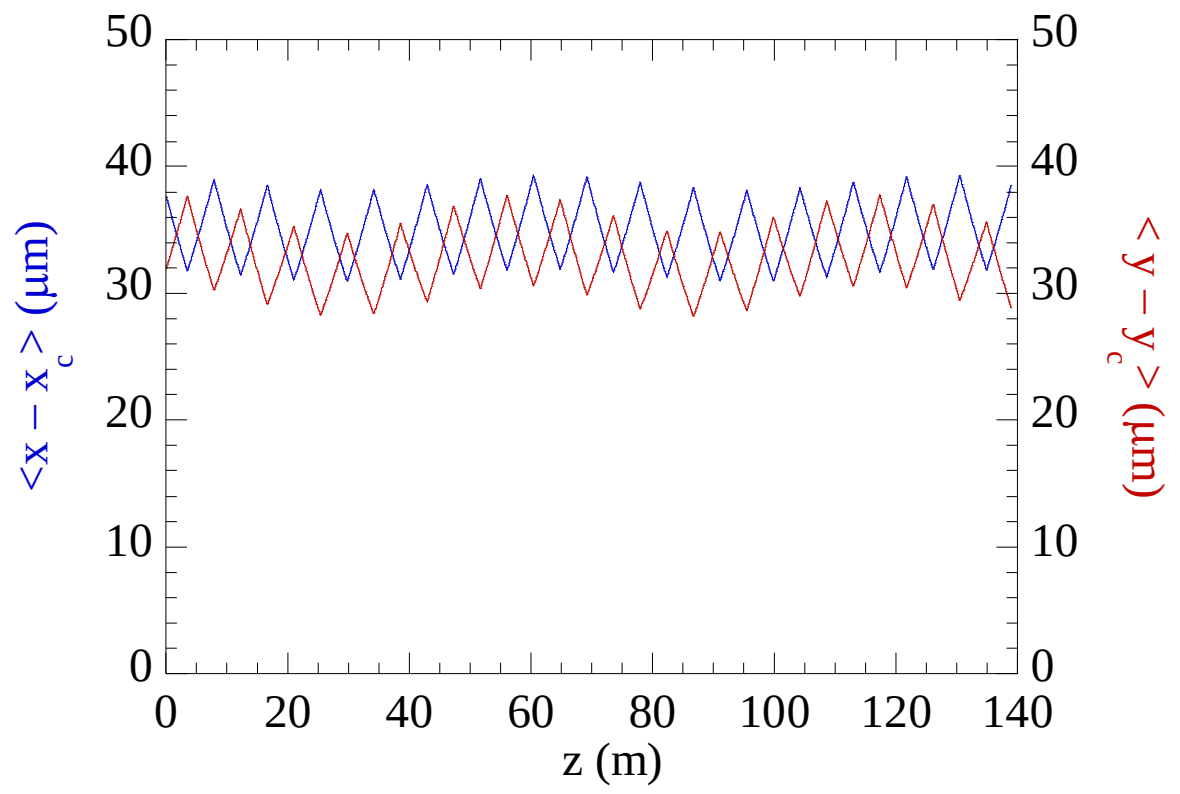

Fig. 4: Propagation of the beam envelope.

We analyse/simulate the case of a fundamental resonance at 3.05 keV (≈ 4.07 Å) which implies an on-axis undulator field strength of 5.61 kG. We assume that the electron beam is characterized by a normalized emittance of 0.45 mm-mrad and a relative energy spread of $1.25 \times 10^{-4}$, which corresponds to the nominal design specification for the LCLS-II This yields a Pierce parameter of $\rho \approx 5.4 \times 10^{-4}$. In order to match this beam into the undulator/FODO line, the initial beam size in the $x$-dimension ($y$-dimension) is 37.87 μm (31.99 μm) with Twiss $\alpha_x = 1.205$ and Twiss $\alpha_y = -0.8656$. Note that this yields Twiss $\beta_x = 24.95$ m and Twiss $\beta_y = 17.80$ m. The propagation of the beam envelopes in $x$ and $y$ corresponding to these parameters are shown in Fig. 4 indicating a good match.

The dimensions of the resonator were fixed based upon estimates of the gain length using the Ming Xie parameterization [47] and MINERVA simulations of the electron beam/undulator line which indicated that approximately 40 – 60 m of undulator would be required for the XFELO to operate as a RAFEL. As such, we fixed the distance, $L$, between the two mirrors framing the undulator as 130 m, which is also the distance separating the two mirrors on the back side of the resonator (elements $C_3$ and $C_4$). In studying the cavity tuning via time-dependent simulations, these two distances are allowed to vary while holding the configurations of the *backscattering* units fixed.

In order to out-couple the x-rays through a transmissive mirror at the wavelength of interest, the diamond crystal would need to be impractically thin (of the order of 5 μm); hence, we consider out-coupling through a hole in the first downstream mirror.

We consider the mirror thickness to be 100 μm and that the out-coupling of the fundamental is through a hole in the first downstream mirror ($C_6$). Due to the high computational requirements of time-dependent simulations, we begin with an optimization of the RAFEL with respect to the hole radius and the undulator length using steady-state (*i.e.*, time-independent) simulations.

The choice of hole radius is important because if the hole is too small then the bulk of the power remains within the resonator while if the hole is too large then the losses become too great and the RAFEL cannot lase. The results for the optimization of the hole radius indicate that the optimum hole radius is 135 μm which allows for 90% out-coupling, where we fixed the undulator line to consist of 11 HXR undulator segments. This is shown in Fig. 5 where we plot the output power as a function of pass number for the optimum hole radius and the variation in the saturated power with the hole radius (inset).

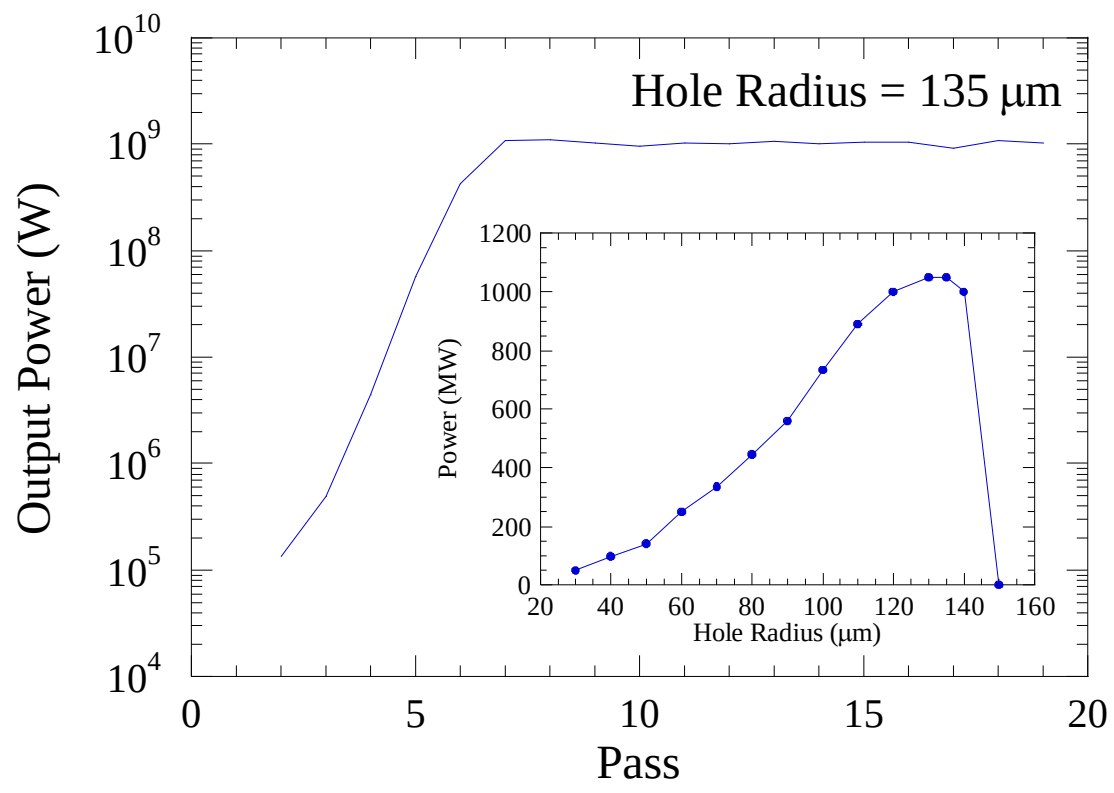

Fig. 5: Output power vs pass for a hole radius of 135 μm and the output power as a function of hole radius (inset).

A local optimization on the undulator length for a hole radius of 135 μm is shown in Fig. 6 where we plot the peak recirculating power (left axis) and the average output power (right axis). The error bars in the figure indicate the level of pass-to-pass fluctuations in the power which is generally smaller than the level of shot-to-shot fluctuations in SASE. Note that while this represents steady-state simulations, the average power is calculated under the assumption of an electron bunch with a flat-top temporal profile having a duration of 24 fs which yields a duty factor of 2.4 × $10^{-8}$. Each point in the figure refers to a given number of HXR undulators ranging from 9 – 13 segments. It is evident from the figure that the optimum length is 47.18 m corresponding to 11 segments.

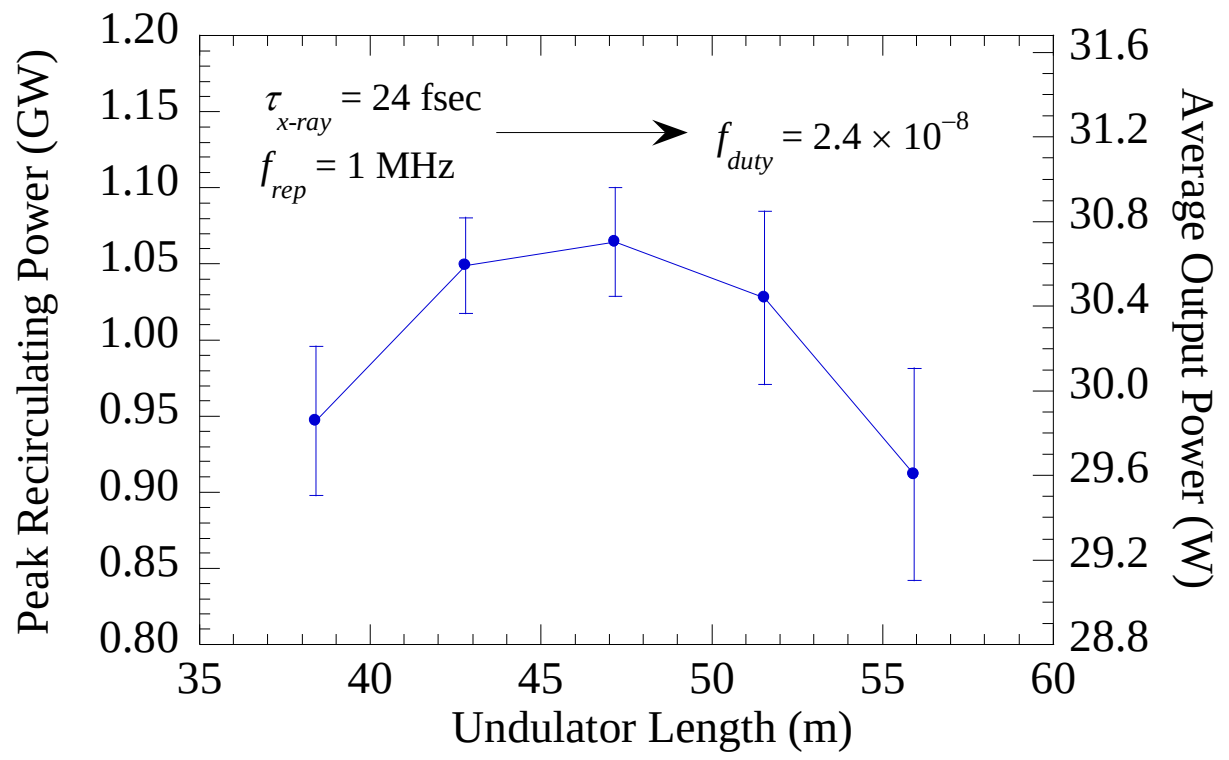

Fig. 6: Optimization on undulator length.

Having found optimum values for the hole radius and the undulator length, we now turn to time-dependent simulations of the RAFEL under the above-mentioned assumption of electron bunches with a flat-top temporal profile having a full width duration of 24 fs and a peak current of 1000 A. This corresponds to a bunch charge of 24 pC.

The first task in running time-dependent oscillator simulations is to determine the detuning curve defining what cavity lengths are synchronized with the repetition rate of the electrons, which is 1 MHz for the LCLS-II. The synchronous cavity length (the so-called zero-detuning length) for the present cavity is $L_{vac} = c/f_{rep}$, where $L_{vac}$ denotes the synchronous, roundtrip cavity length for the vacuum resonator, $f_{rep}$ is the repetition rate and $c$ is the speed of light *in vacuo*. For the case under consideration $L_{vac}$ = 299.7924580 m. The range of cavity lengths that can result in synchronism between the optical pulse and the electron bunches is affected by the electron bunch length, which is about 7.2 μm for the present configuration. Thus, it is expected that some synchronism may be expected for cavity lengths in the range of $L_{vac}$ – 7.2 μm < $L_{cav}$ < $L_{vac}$ + 7.2 μm, where $L_{cav}$ denotes the total roundtrip length of the cavity.

Turning to multi-pass, time-dependent simulations, we begin by considering a hole radius of 135 μm and an rms energy spread of 0.0125% (the nominal value for the LCLS-II). The temporal profile was taken to be flat-top with a full width of 24 fs yielding a bunch charge of 24 pC. This is within the expected range for the LCLS-II but is not the maximum possible bunch charge; hence, further simulations using a higher bunch charge may yield still higher output power than found at the present time. We consider start-up from noise on the electron beam on the first pass (with noise included in the simulations for each successive pass as well) and since the RAFEL employs a high-gain undulator line the pulse energy after the first pass reaches about 5 nJ, and subsequent growth is rapid despite an out-coupling from the first downstream mirror of about 90% of the incident pulse. The detuning curve is shown in Fig. 7 and we note that gain is observed for both positive and negative detuning over ranges of about ±7.2 μm from perfect synchronism, which corresponds to the duration of the electron bunch. Typically, saturation is achieved after about 15 – 25 passes. Note that once the cavity is detuned by more than the electron bunch length (either positive or negative) the RAFEL fails to lase. As shown in the figure, since the single-pass gain is high in a RAFEL, this transition to complete desynchronization occurs rapidly. We note that the output pulse energy at the peak of the detuning curve is about 21.2 μJ and the average output power is about 21.2 W for 3.05 keV photons. Given an assumed bunch duration of 24 fs and a repetition rate of 1 MHz, this implies that the output power per pulse would be about 880 MW and the long-term average output power of about 21.2 W.

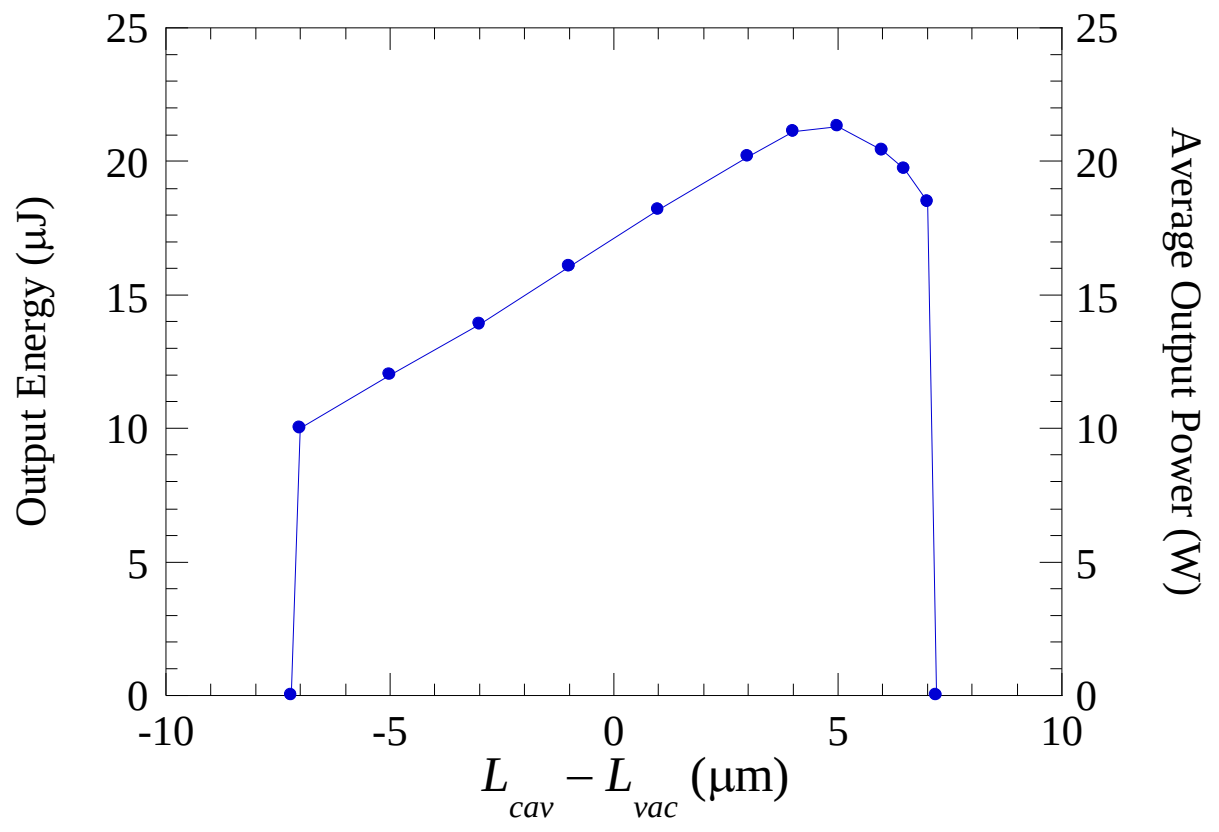

Fig. 7: Cavity detuning curve.

The evolution of the output energy at the fundamental and the 3rd harmonic, and the spectral linewidth of the fundamental, vs pass are shown in Fig. 8 for a detuning of 5 μm which is close to the peak in the detuning curve (Fig. 7). While it is not evident in the logarithmic scale of the output energy, the rms fluctuation in the energy from pass to pass is of the order of about 3 percent, which is lower than the shot-to-shot fluctuations expected from pure SASE. At least as important as the output power is that the linewidth contracts substantially during the exponential growth phase and remains constant through saturation. Starting with a linewidth of about $3.7 \times 10^{-4}$ after the first pass corresponding to SASE, the linewidth contracts to about $6.0 \times 10^{-5}$ at saturation. The SASE linewidth after the first pass through the undulator is slightly smaller than the predicted SASE linewidth based on 1-D theory [48] which is approximately $4.5 \times 10^{-4}$. Hence, the RAFEL is expected to have both high average power and a stable narrow linewidth.

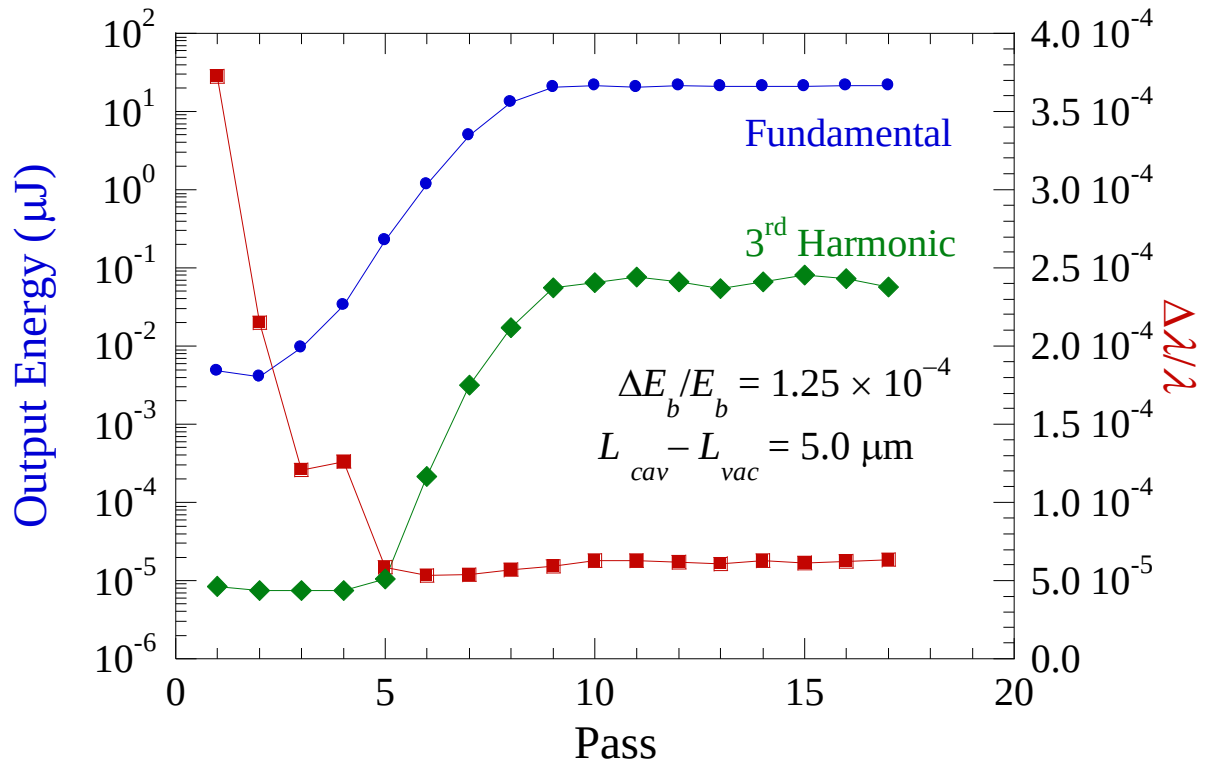

Fig. 8: Evolution of the output pulse energy (left axis) at the fundamental (blue) and the third harmonic(green), as well as the linewidth (red, right axis).

The 3rd harmonic grows parasitically from high powers/pulse energies at the fundamental in a single pass through the undulator [49] and has been shown to reach output intensities of 0.1% that of the fundamental in a variety of FEL configurations. Indeed, this is what is found in the RAFEL simulations. As shown in Fig. 8, the 3rd harmonic intensity remains small until the fundamental pulse energy reaches about 1 μJ after which it grows rapidly and saturates after about 12 passes. This is close to the point at which the fundamental saturates as well. The saturated pulse energies at the 3rd harmonic reach about 0.067 μJ. Given a repetition rate of 1 MHz, this corresponds to a long-term average power of 67 mW.

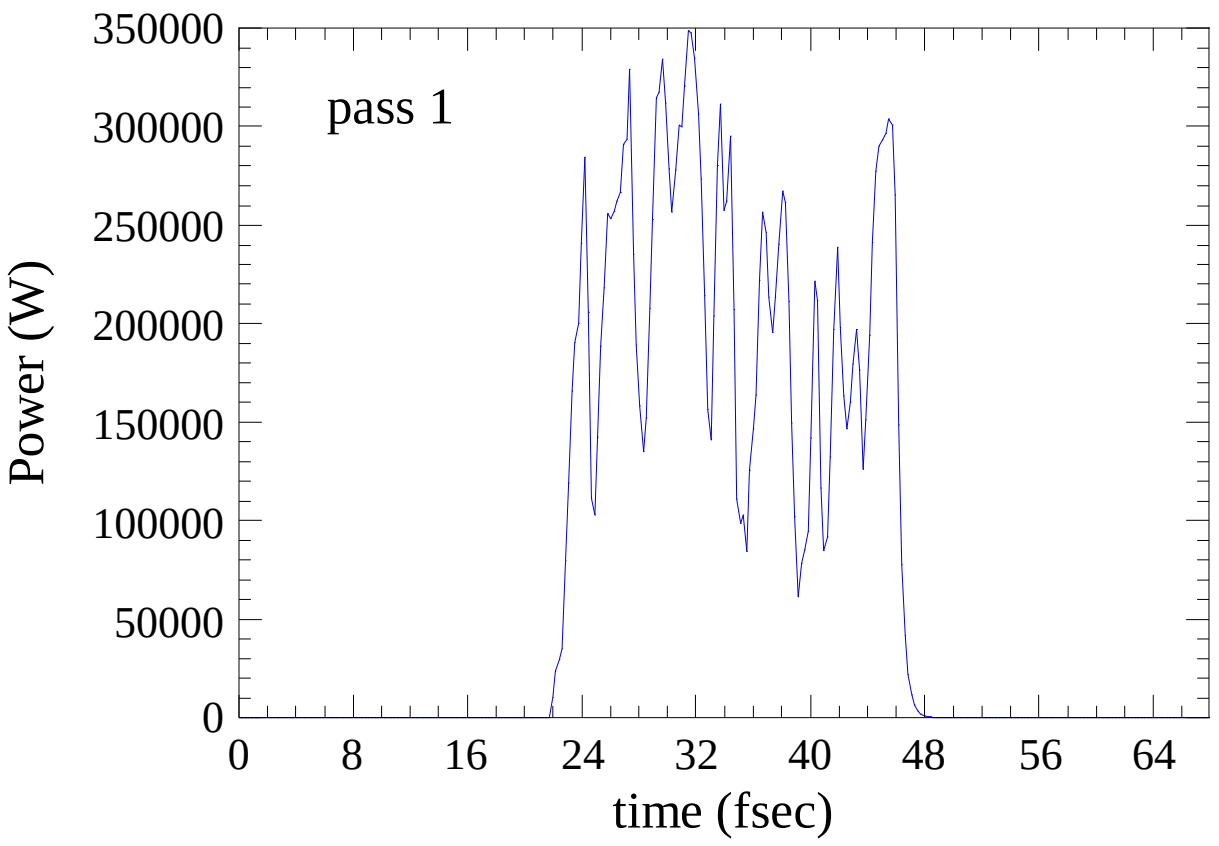

Fig. 9: Temporal profile of the optical pulse at the undulator exit after the first pass.

The reduction in the linewidth after saturation shown in Fig. 8 indicates that a substantial level of longitudinal coherence has been achieved in the saturated regime. The RAFEL starts from shot noise on the beam during the first pass through the undulator, and longitudinal coherence develops over the subsequent passes. Hence, we expect that the temporal profile of the optical field will exhibit the typical spiky structure associated with pure SASE after the first pass through the undulator and this is, indeed, what we find as shown in Fig. 9. The number of spikes expected, $N_{spikes}$, is given approximately by $N_{spikes} \approx l_b/(2\pi l_c)$, where $l_b$ is the rms bunch length and $l_c$ is the coherence length. For the present case, $l_b \approx 7.2$ μm and $l_c \approx 60$ nm; hence, we expect that $N_{spikes} \approx 19$. The temporal profile of the optical pulse at the undulator exit after the first pass is shown in Fig. 9, where we observe about 14 spikes which is in reasonable agreement with the expectation. Note that the time axis encompasses the time window used in the simulation.

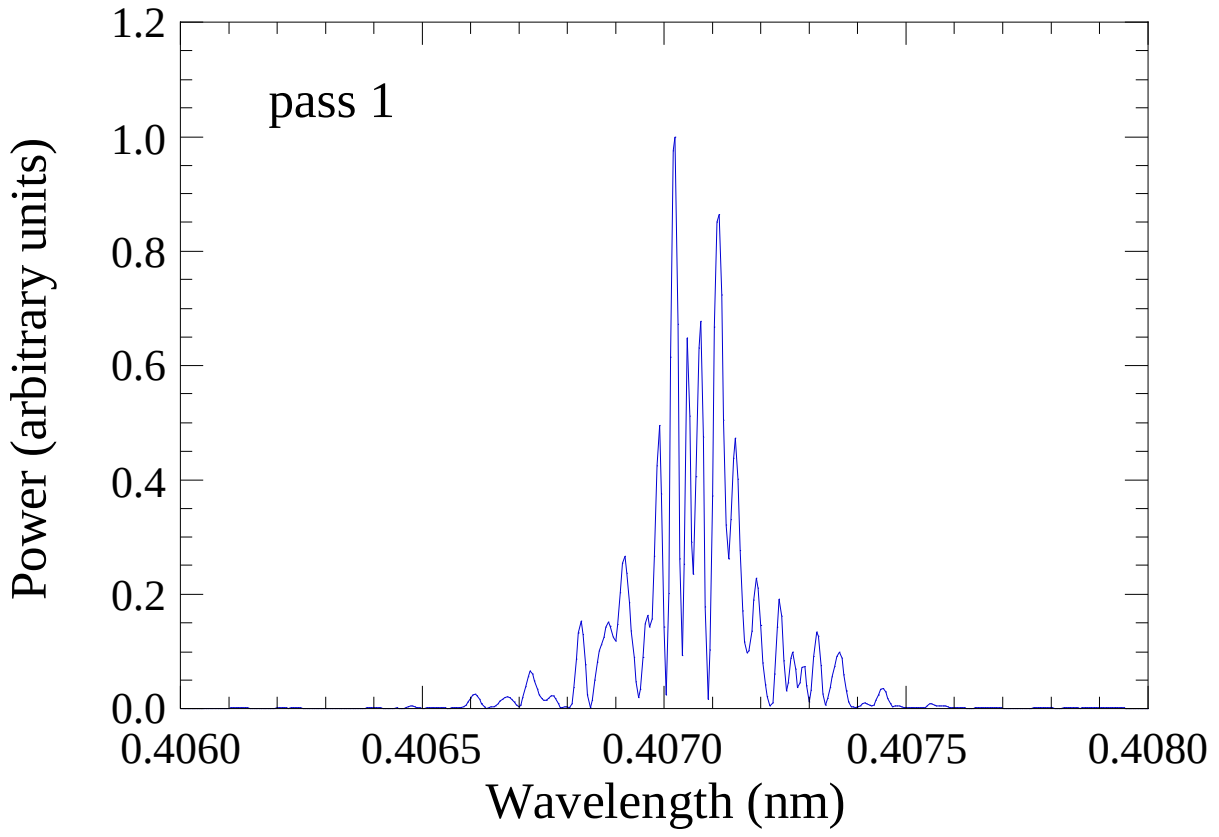

Fig. 10: Spectrum at the undulator exit after the first pass.

As indicated in Fig. 8, the linewidth after the first pass is of the order of $4.3 \times 10^{-4}$ which is relatively broad and corresponds to the interaction due to pure SASE. This is reflected in the output spectrum from the undulator after the first pass which is shown in Fig. 10.

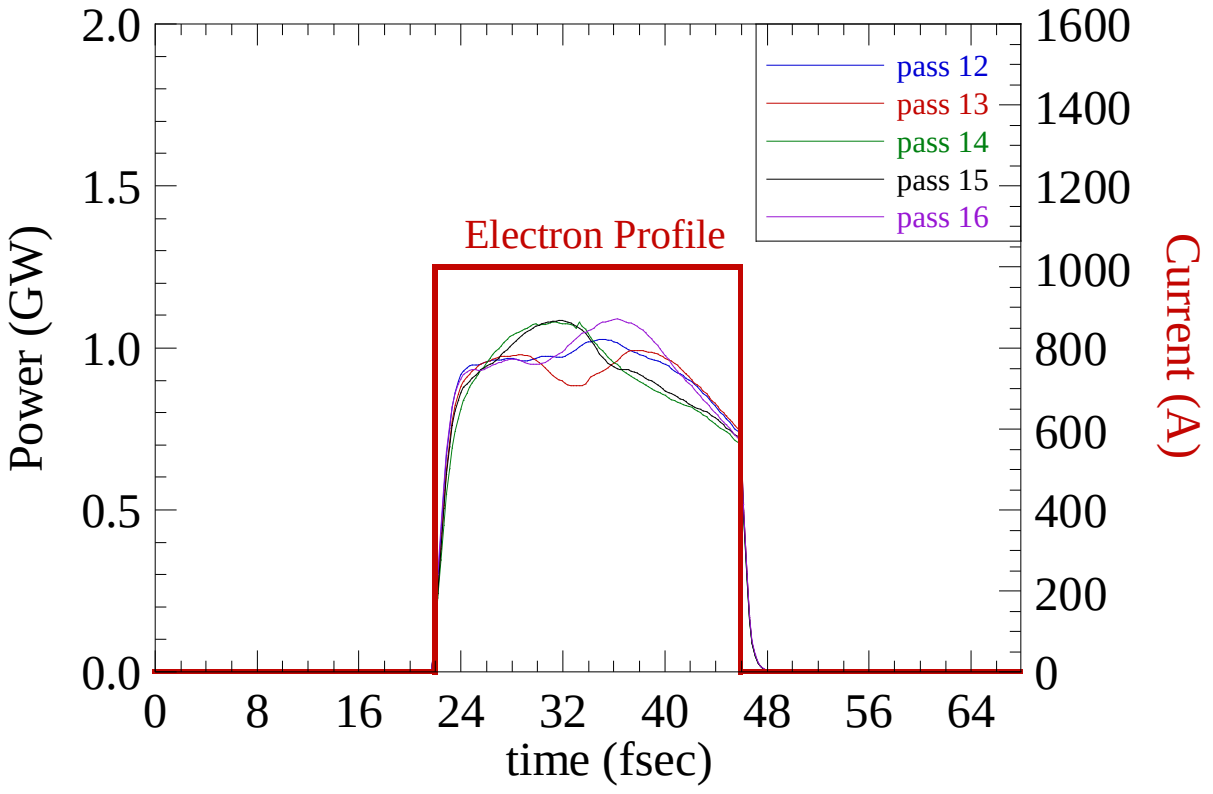

Fig. 11: Temporal profiles of the x-ray pulse at the undulator exit for different passes in saturation (left axis) and the electron bunch profile (right axis).

The spectral narrowing that is associated with the development of longitudinal coherence as the interaction approaches saturation results in a smoothing of the temporal profile. This is illustrated in Fig. 11, where we plot the temporal profiles of the optical field at the undulator exit corresponding to passes 12 – 16 which are after saturation has been achieved (left axis). As shown in the figure, the temporal pulse shapes from pass-to-pass are relatively stable and exhibit a smooth plateau with a width of about 23 – 24 fs which corresponds to, and overlaps, the flat-top profile of the electron bunches which is shown on the right axis. Significantly, the smoothness of the profiles corresponds with the narrow linewidth and contrasts sharply with the SASE output after the first pass through the undulator (see Fig. 9). Both the pass-to-pass stability and smoothness of the output pulses contrast markedly with the large shot-to-shot fluctuations and the spikiness expected from the output pulses in pure SASE.

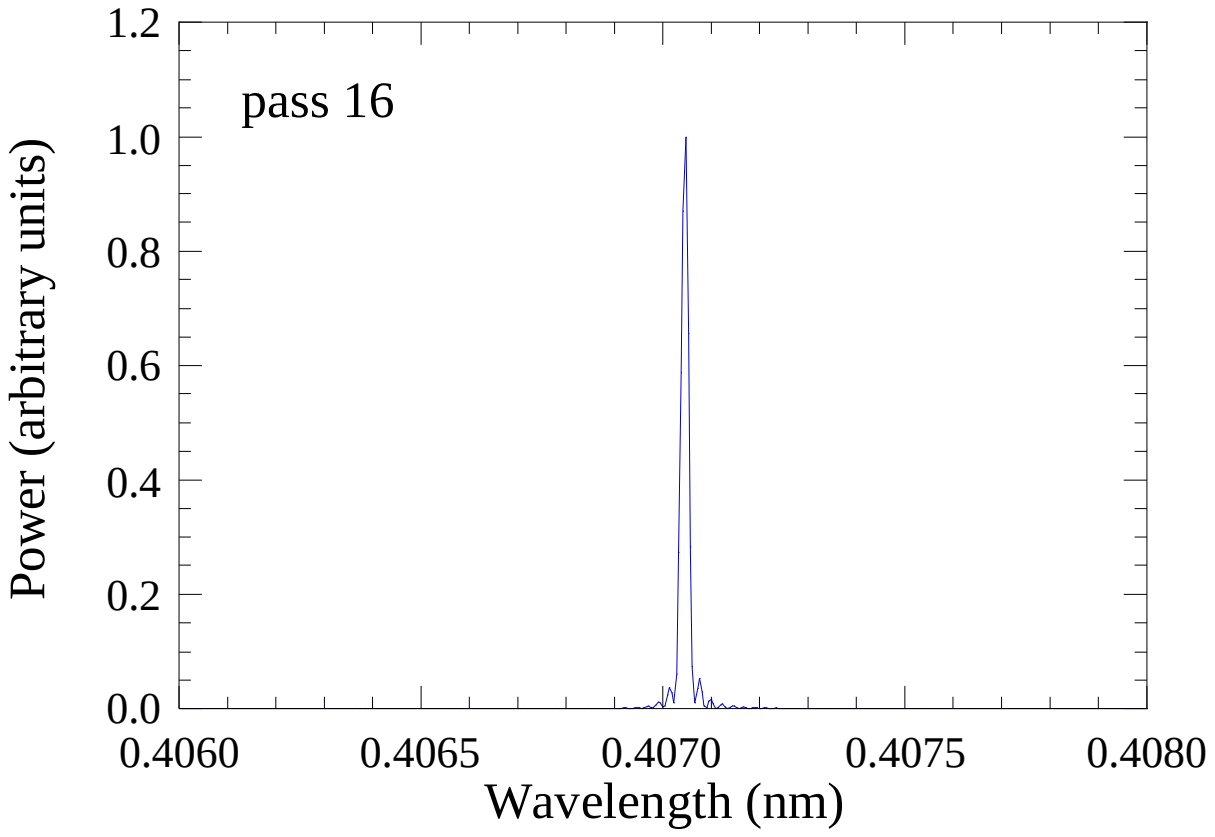

Fig. 12: Spectrum at the undulator exit after pass 16.

The narrow relative linewidth in this regime of about $7.3 \times 10^{-5}$ at the undulator exit, as shown in Fig. 12 after pass 16, as well as the smooth temporal profiles shown in Fig. 11, are associated with longitudinal coherence after saturation is achieved. Note the comparison with the spectrum after the first pass which corresponds to pure SASE.

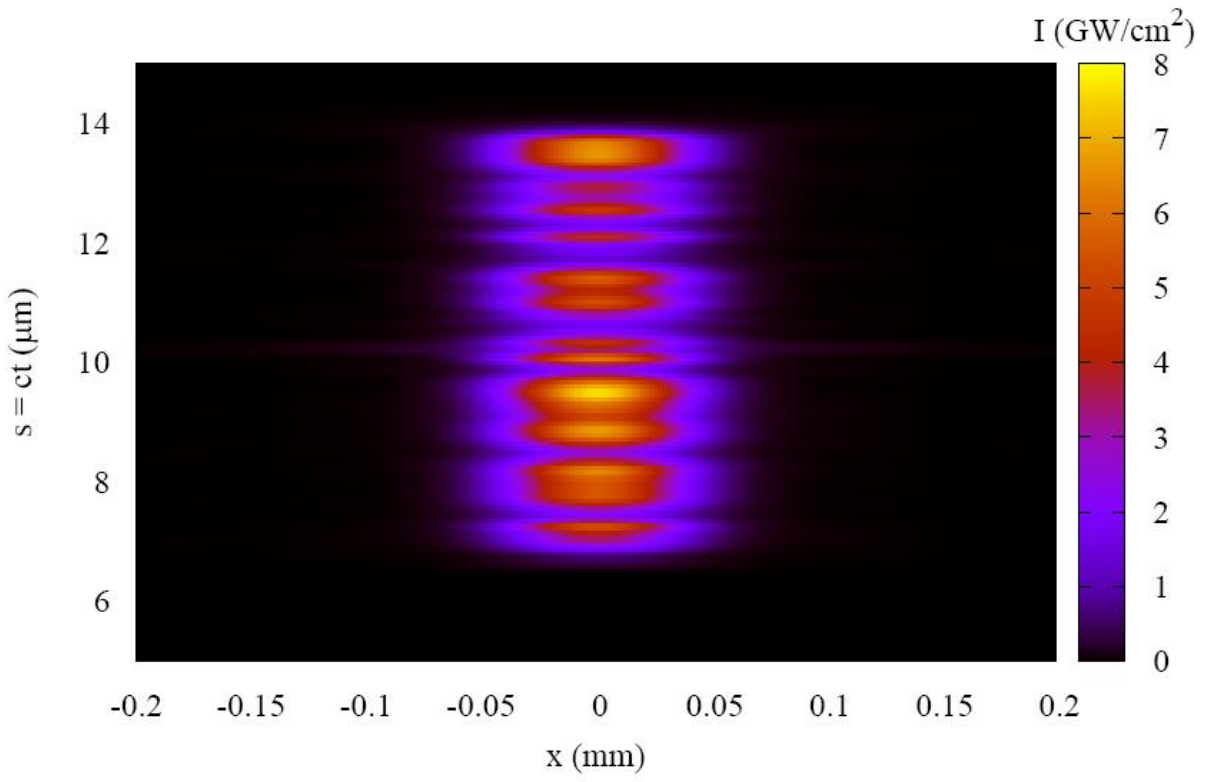

Fig. 13: Contour plot showing the temporal and transverse mode content at the undulator exit on the first pass.

The transverse mode structure is shown in Figs. 13 and 14 which shows contour plots of the profile at the undulator exit after the first pass and the 16th pass (which is deep into saturation), where the longitudinal structure is indicated by the vertical axis and the transverse mode structure is along the horizontal axis. Note that the horizontal axis corresponds to the wiggle-plane. The spiky nature of SASE after the first pass through the undulators is clearly shown in Fig. 13 which also shows that the mode is well-localized transversely within the electron bunch. After saturation, however, the output pulse is spatially and temporally smooth as shown in Fig. 14, which is of great benefit for numerous applications.

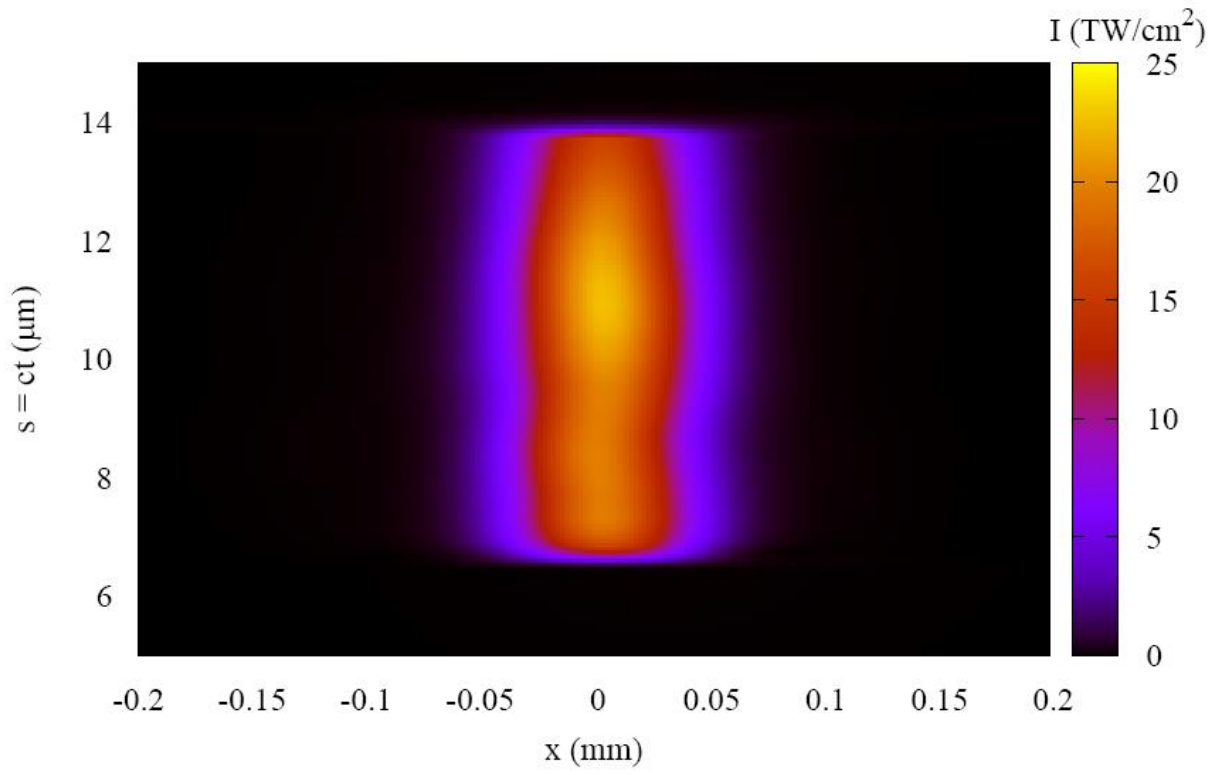

Fig. 14: Contour plot showing the temporal and transverse mode content at the undulator exit on the 16th pass which is deep into saturation.

## V. SUMMARY AND CONCLUSION

In this paper, we describe a design for an x-ray RAFEL using a six-mirror resonator composed of diamond crystal Bragg reflectors with pinhole out-coupling and the LCLS-II beam and undulator line. Using the 4.0 GeV beam available from the LCLS-II and the HXR undulator (2.6 cm period, and 5.6 kG on-axis field), we considered the generation of 3.05 keV (4.07 Å wavelength) x-rays. The x-rays were out-coupled through a hole in the first downstream mirror whose optimum radius was found to be 135 μm and through which more than 90% of the x-ray energy was out-coupled from the resonator. Under the assumption of 24 pC electron bunches and the nominal energy spread of $1.24 \times 10^{-4}$, simulations indicate that the peak out-coupled pulse energy was about 22 μJ corresponding to an average output power of 22 W. Furthermore, the spectral width was found to decrease markedly as saturation is reached. Third harmonic output was also found to be significant and to reach average output powers of about 60 mW. Further, the output pulse shapes closely correspond to the temporal profile of the electron bunches and are relatively smooth and stable from pass-to-pass. As such, we conclude that an x-ray RAFEL may constitute an important alternative to pure SASE XFELs.

The present state-of-the-art in the production of diamond crystals can provide nearly flawless diamond crystals featuring close to 99% Bragg reflectivity of hard x-rays [22,23]. A new aspect of the present study is, however, the proposed out-coupling through a pinhole in one of the diamond crystal mirrors. In the present case, we considered a pinhole with a diameter of 270 μm. Diamond is one of the hardest materials and is chemically inert. Mechanical or chemical (including plasma etching) machining techniques are slow and inefficient. An important issue, therefore, is whether diamond mirrors with pinholes can be manufactured with the high crystalline perfection necessary to ensure high Bragg reflectivity. It is our opinion that such pinhole mirrors can be manufactured by using an alternative method of machining.

Laser ablation seems to be an appropriate method for the fabrication of pinhole diamond crystal mirrors, and offer a way to control the crystal hole form fidelity with laser beams. Use of ultra-short (picosecond or femtosecond) laser pulses are essential, as they can mill materials with a small amount of heating and residual damage. The recently demonstrated feasibility of manufacturing diamond parabolic lenses [50,51] and diamond drumhead crystals [26] by laser milling suggests that this technique could also be successful in its application to the fabrication of diamond pinhole crystal mirrors.

## ACKNOWLEDGEMENTS

This research was supported under DOE Contract DE-SC0018539. The research used resources of the Argonne Leadership Computing Facility, which is a DOE Office of Science User Facility supported under contract DE-AC02-06CH11357. We also thank the University of New Mexico Center for Advanced Research Computing, supported in part by the National Science Foundation, for providing high performance computing resources used for this work.